\documentclass{ws-ijmpa}
\usepackage{amsmath}
\usepackage{amssymb}
\usepackage{epsfig}

\newsavebox{\zzzbar}
\sbox{\zzzbar}
  {\setlength{\unitlength}{0.9em}
  \begin{picture}(0.6,0.7)
  \thinlines
  \put(0,0){\line(1,0){0.6}}
  \put(0,0.75){\line(1,0){0.575}}
  \multiput(0,0)(0.0125,0.025){30}{\rule{0.3pt}{0.3pt}}
  \multiput(0.2,0)(0.0125,0.025){30}{\rule{0.3pt}{0.3pt}}
  \put(0,0.75){\line(0,-1){0.15}}
  \put(0.015,0.75){\line(0,-1){0.1}}
  \put(0.03,0.75){\line(0,-1){0.075}}
  \put(0.045,0.75){\line(0,-1){0.05}}
  \put(0.05,0.75){\line(0,-1){0.025}}
  \put(0.6,0){\line(0,1){0.15}}
  \put(0.585,0){\line(0,1){0.1}}
  \put(0.57,0){\line(0,1){0.075}}
  \put(0.555,0){\line(0,1){0.05}}
  \put(0.55,0){\line(0,1){0.025}}
  \end{picture}}

\newsavebox{\uuunit}
\sbox{\uuunit}
    {\setlength{\unitlength}{0.825em}
     \begin{picture}(0.6,0.7)
        \thinlines
        \put(0,0){\line(1,0){0.5}}
        \put(0.15,0){\line(0,1){0.7}}
        \put(0.35,0){\line(0,1){0.8}}
       \multiput(0.3,0.8)(-0.04,-0.02){12}{\rule{0.5pt}{0.5pt}}
     \end {picture}}

\def\nc{non-commutative}
\newcommand{\bea}{\begin{eqnarray}}
\newcommand{\eea}{\end{eqnarray}}
\newcommand{\be}{\begin{equation}}
\newcommand{\ee}{\end{equation}}
\def\q{&=&}
\def\ft{fuzzy torus}
\def\12{\frac{1}{2}}
\def\udag{U^{\dag}}
\def\vdag{V^{\dag}}
\def\xdag{X^{\dag}}
\def\ydag{Y^{\dag}}
\def\zdag{Z^{\dag}}
\def\ep{e^{i \theta}}
\def\epp{e^{2i \theta}}
\def\epn{e^{i(N-1) \theta}}
\def\exm{e^{-i \theta}}
\def\x{{\cal{X}}}
\def\y{{\cal{Y}}}
\def\xd{{\cal{X}}^{\dag}}
\def\yd{{\cal{Y}}^{\dag}}
\def\wl{Wilson line}
\def\da{^{\dag}}
\def\N{{\cal N}}

\begin{document}

\markboth{D. Bigatti}
{String Theory and the Fuzzy Torus}

%
\catchline{}{}{}{}{}
%

\title{STRING THEORY AND THE FUZZY TORUS}

\author{\footnotesize DANIELA BIGATTI}

\address{Department of Particle Physics, Weizmann Institute\\
Rehovot 76100, Israel
}

\maketitle

\pub{Received (Day Month Year)}{Revised (Day Month Year)}

\begin{abstract}
We outline a brief description of non commutative geometry and present some 
applications in string theory. We use the fuzzy torus as our guiding 
example.

\end{abstract}

\section{Introduction}
Geometry, the science of properties of spaces, has been, along the centuries, referring to different definitions of its domain of interest. As a general fact, the tendence has been to widen the scope of geometry and the concept of "admissible spaces", along with an abstraction process that shifted the focus from the properties of spaces (as experienced by an "external" observer living in some embedding geometrical entity, or by an "intrinsic" observer living in the space itself) to the relations among entities belonging to the space ("geometrical objects"). At the same time, the features of spaces have evolved from being very concrete and almost physical, to a level of abstraction which essentially rephrases them in analitical terms. Geometry did not certainly become less interesting along this path, which has also been giving birth to many interesting and fruitful branches of mathematics - and noncommutative geometry is just one of them. We will here aim to sketch a brief outline of non commutative geometry and its possible applications, and discuss some examples of string theory models which enjoy a non commutative description. 

\section{Non-commutative geometry}
Both physics and mathematics, and in particular geometry which specifically is devoted to the study of spaces, often deal with the task of investigating objects whose direct analysis is overly difficult, exceedinly cumbersome or impossible
tout court. 

Let us consider, for example, a topological space. Given a generic set $X$ endowed with a topology, $\mathcal{M}$, handling directly its topology would involve not only heavy operations overs sets that we are not familiar with, but would also require the handling of an enormeous number of subsets --- they may be easily as many as $\mathcal{P}(X)$. So such an approach is reserverd to the situation in which either the topology is naturaly ``easy'' --- but, often, uninteresting --- or when it is possible to isolate a basis foe the topology, that is if we can completely derive it out of a small number of open sets. 

The standard procedure to study a topological space is to realize that it is the natural habitat to grow the notion of a continuos function. We consider the function $f:\,X\to\mathbb{R}$, and exploit the ease to do computations with numbers and the fact that the numerical fields have an obvious notion of ``neighbour'' among the complexity of their structures. We will say that $f$ is continuous if the counterimage of an open set of $\mathbb{R}$ is still an open set for the topology $\mathbb{M}$ chosen on $X$. And we can then proceed to analyze the class of continuous functions over $X$, which enjoys naturally the structures of a commutative algebra, because it inherits them from the structures of real numbers. It is enough to define the sum and the product of functions as the function whose value at each point is the sum or product of the numerical values of the functions on each point: $(f {\mathbf +}g) (x) := f(x) +
g(x)$, $(f {\mathbf \cdot}g) (x) := f(x) \cdot g(x)$.
Since this is obtained from the addition and multiplication of $\mathbb R$ or
$\mathbb C$ only, we  remark that the same construction can be carried out 
for, as
an example, matrix valued continuous functions, the only difference being the
loss of commutativity.  
\par The idea of studying a topological space through
the algebra of its continuous functions is the central idea of algebraic
topology. Likewise, algebraic geometry studies characteristics  of spaces by
means, for example, of their algebra of rational functions. In this framework
it is  very natural to ask what happens if we replace the algebra of ordinary 
functions with some non-commutative analogue, both to extend the tools to new 
and previously intractable contexts (and sometimes this 
replacement is unavoidable) and to study with more refined probes the 
``classical spaces'' (which leads sometimes to new and surprising results). 

\subsection{Quotient spaces}

A class of typically intractable (and typically very
interesting) objects is reached by means of  a quotient space construction,
that is, by considering a set endowed with an equivalence  relation fulfilling
reflexivity, symmetry and transitivity axioms and identifying the elements
which  are equivalent with respect to such a relation. In the following we
will find particularly useful  the ``graph'' picture of an equivalence 
relation: if we consider the Cartesian product of two copies of the set, we can
assign the  equivalence relation as a subset of the Cartesian product (the one
formed by the couples  satisfying the relation). The construction gives
interesting results already if we consider a set  (possibly with an operation),
but of course is much richer if we act over a set with structure.
\par  Let's clarify our statement with an example \cite{Connes}. Consider
the flat square torus, that is, $[0,1] \times [0,1]$ with the opposite sides 
ordinately glued. This is clearly a well-behaved space and, undoubtedly, a
compact one\footnote{ We say that a topological 
space with a given topology is compact if any open covering  of the space
(i.~e.~any family of open sets whose union is the space itself)  admits a
finite subcovering (i.~e.~a finite subfamily of the above which is  still a
covering). We say thet a topological space $(X, {\cal U})$ is locally  compact
if for all $x \in X$ and for all $U \in {\cal U}$, $x \in U$, there  exist a
compact set $W$ such that $W \subset U$.  It is, of course, useful to refer to
a compactness notion also  when we have more structures than just the one of a 
topological space (for example, differentiability).}.  Let's introduce an equivalence relation which identifies the
points of the lines parallel  to $y= \sqrt{2} x$, that is, we ``foliate'' the
space into leaves parametrized by the intercept. Since  $\sqrt{2}$ is
irrational, though, any such leaf fills the torus in a dense way (that is,
given a leaf  and a point of the torus, the leaf is found to be arbitrarily
close to the point). If we try to study the  quotient space and to introduce in
it a topology, we will find that ``anything is close to anything'',  that is,
the only possible topology contains as open sets only the whole space and the
empty  set. It is hopeless to try to give the quotient space an interesting
topology  based on our notion of ``neighborhood'' of the parent space.  
\par It is, in particular, hopeless for all practical purposes to give the 
space the standard notion of topology inherited by  the quotient operation, 
which we shortly describe. If we have a space $A \equiv B / \sim$,  there is a 
natural projection map  \begin{eqnarray}  \begin{array}{c}  p \: : 
\:\:\:\:\:\:\:  B \longrightarrow A  \\  p \: : \:\:\:\:\:\:\:  x \longmapsto 
[x]  \end{array}
\end{eqnarray} 
which sends $x \in B$ in its equivalence class. The inherited
topology on $A$  would be the one  whose open sets are the sets whose
counterimages are open sets in $B$.  
\par We want an interesting topology,
richer than $\{  \emptyset , X \}$ ,  and, moreover, we would like the
topological space so obtained to enjoy local  compactness. The reason why we
make the effort is that the dull topology  $\{  \emptyset , X \}$ treats the
space, from the point of view of continuous  functions which will be our probe,
as the space consisting of only one point;  all the possible subtleties of our
environment will be lost. Local  compactness is a slightly more technical tool,
but we can imagine, both from  the physical and the mathematical point of view,
why it is so useful.  Each time we have a nontrivial bundle (and we will have
plenty of them in  the following) we usually define them not globally, but on
neighbourhoods.  Since these ``patches'' will in general intersect, we need a
machinery  to enforce agreement of alternative descriptions. Local compactness
(and  similar tools) ensure us that ``the number of possible alternative 
descriptions will never get out of control''. We shall see how to achieve the 
notable result of introducing in ``weird'' spaces a rich enough and  even
locally compact topology. An example of the process is presented in the  next
section. 

\subsection{A typical example: the space of Penrose tilings}
We are going to
discuss a situation which embodies most of the characteristic features both of 
the problems which non-commutative geometry makes tractable and of the procedure
which  allows their handling\cite{tasnadi}. 
\par Penrose was able to build tilings of the
plane having a 5-fold symmetry axis; this is not  possible by means of periodic
tilings with all equal tiles, as it is known  since a long time.
They are composed (see Fig.~1) of two types of tiles:  ``darts'' and 
``kites'', with
the condition that every vertex has matching  colors. A striking characteristic
of the Penrose tilings is that any finite  pattern occurs (infinitely many
times, by the way) in any other Penrose  tiling. So, if we call identical two
tilings which are  carried into each other by an isometry of the plane (this is
a sensible  definition since none of the tilings is periodic), it is never
possible to decide locally which tiling is which. However the distinct Penrose tilings 
are an uncountable infinity.

\begin{figure}
\centerline{\psfig{file=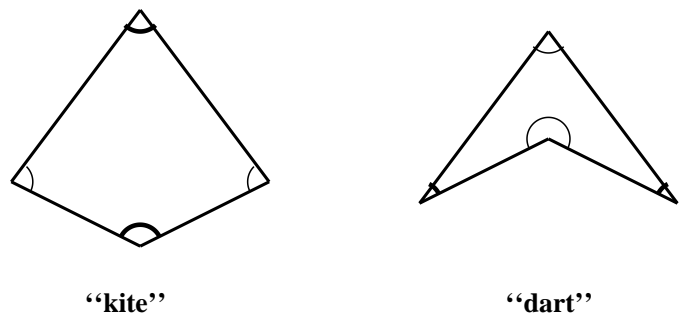,width=10cm}}
\vspace*{8pt}
\caption{Dart and Kites.}
\end{figure}

Yet one can prove that there exist numbers (that is integer numbers) that label (some of)
the subtly different elements of the space --- and it would be quite unpleasant to dismiss
such a fact. After having shown that the space of distinct Penrose tilings can be described as
$ X = K / {\cal R}$, where $K$ is compact and ${\cal R}$ is some equivalence relation\footnote{This point
could be quite confusing to the
careful reader: since the ${\cal R}$ equivalence
classes contain a denumerable infinity of elements, while $K$ has the
cardinality of continuum, obviously the quotient space has more than one (and actually
an awful lot of) elements. What we want to do is not a counting,  but to make
sure that the tilings are different in an {\it interesting} sense.}, the path
of non-commutative geometry is precisely to show (1) that the attempt of distinguishing the 
tilings by means of  an algebra of operator-valued functions is successful, and (2) that
it is actually sensible to say that there are different Penrose tilings, since
topological invariants can be built and used to label the tilings. For a very elementary
summary, see \cite{danprimer}.

\section{Yang-Mills Theory on the Fuzzy Torus}
In the next two sections we will concentrate on the simplest, yet physically non-trivial, example
of a non-commutative geometry: the fuzzy torus. We will define the notion of a bundle
over such geometry, which leads to what physicists call a gauge theory. We will also explain
how string theory provides a setting in which such structures arise naturally 
\cite{PWW,PW,KO,PMH,LLS,bmz,mz}.

\subsection{Foliated torus and the non-commuting $U$, $V$ algebra}
We would first like to clarify the relation between the so-called
non-commuting torus geometry and the picture of the torus foliation, 
that is required in order to describe the non-commutative torus as a quotient, as outlined 
in the previous section. We will consider the
situation where the slope of the leaf is irrational (that is, the leaf is infinite).
Let us recall the main features of the two approaches.
\par In the non-commuting torus representation, one generalizes the toroidal
geometry by supposing that the two coordinates are not ordinary variables,
but satisfy the commutation relations
\begin{eqnarray} \displaystyle{
[p, q] = i\,\theta
} \end{eqnarray}
This equation cannot of course be satisfied by finite dimensional matrices (no two
finite dimensional matrices can have a commutator with a non-vanishing trace), but 
can be satisfied by a pair of operators, $q$ and $p$, on a Hilbert space.
A more convenient labeling in order to provide an useful representation
in terms of finite dimensional matrices is
\begin{eqnarray}
\label{commrel}
U = e^{ip}\,,\quad V=e^{iq}\quad \Rightarrow \quad UV = e^{i\,\theta} VU
\end{eqnarray}
In the case $\theta= 2\pi/N$, one can write down explicit $N\times N$ matrices that satisfy the 
above relation, and will turn out to be very useful in the following. Let's have as $U$ the 
shift matrix 
\begin{eqnarray}\displaystyle{ 
U= \left( 
\begin{array}{cccccc}
0 & 1 & 0 & 0 & 0 & 0 \\
0 & 0 & 1 & 0 & 0 & 0 \\
0 & 0 & 0 & 1 & 0 & 0 \\
0 & 0 & 0 & 0 & 1 & 0 \\
0 & 0 & 0 & 0 & 0 & 1 \\
1 & 0 & 0 & 0 & 0 & 0 \\
\label{shift}
\end{array}
\right) 
} \end{eqnarray} 
and as $V$ the matrix of phases 
\begin{eqnarray} \displaystyle{ 
V= \left( 
\begin{array}{cccccc}
1 & 0 & 0 & 0 & 0 & 0 \\
0 & e^{2 \pi i / N} & 0 & 0 & 0 & 0 \\
0 & 0 & e^{4 \pi i / N} & 0 & 0 & 0 \\
0 & 0 & 0 & e^{6 \pi i / N} & 0 & 0 \\
0 & 0 & 0 & 0 & e^{8 \pi i / N} & 0 \\
0 & 0 & 0 & 0 & 0 & \ddots \\
\label{phases}
\end{array}
\right) 
} \end{eqnarray}  
It is straightforward to check that (\ref{shift}), (\ref{phases}) satisfy
\bea
U^{\dag}U&=&V^{\dag}V =1 \cr
U^N&=&V^N =1 \cr
UV &=& VU e^{i \theta}
\label{1.1}
\eea
with $\theta = {2 \pi} /N $. 
\par Functions on the \ft \ are defined by the \nc \ analogue of a Fourier
series: for any  $N \times N$ matrix $\phi(U,V)$, the expansion
\be
\phi(U,V)=\sum_{n,m=0}^{N-1} c_{mn}U^nV^m
\ee
associates to it the function of  two periodic variables whose Fourier modes are $c_{n,m}$.
It is convenient to define
\bea
U_{mn}&=&e^{-imn\theta} U^mV^n \cr
\phi_{mn} &=&e^{imn\theta}c_{mn}\cr
\phi(U,V)&=&\sum_{n,m=0}^{N-1}\phi_{mn}U_{mn}
\label{1.7}
\eea
Note that the $U_{mn} $ satisfy
\be
U_{mn}U_{rs}=\exp{\12 i{\theta }(ms-nr)} \equiv U_{mn}*U_{rs}
\label{1.8}
\ee
Equation (\ref{1.8}) defines the star-product on the \ft .
\par Let us switch to the foliated torus picture. 
Consider the square unit torus, that is, $[0,1] \times [0,1]$ with 
the opposite sides ordinately glued. The foliation
of ${\mathbb R}^2$ generated by a family of lines of fixed slope:
\begin{eqnarray} \displaystyle{
y=ax + b \: \: \: \: \: \: \: \: \: \: \: \: \: \: \: \:\: \: \: \: \: (a \: 
{\mathrm{fixed}})
} \end{eqnarray}
induces a corresponding foliation of the torus. Identifying the points belonging to any such leaf
leads to the fuzzy torus. 
\par A leaf of the foliation is parametrized by the value of $b$, but in a 
very redundant way. Actually two values of the intercept correspond to the 
same leaf provided that 
\begin{eqnarray} \displaystyle{
b'= b+an \: \: \: \: \: \: \: \: \: \: \: \: \: \: \: \: \: \: \: \: \: 
n \in {\mathbb Z}
} \end{eqnarray}
Now let's introduce functions of two variables (one of which is an integer): 
$F(b, n)$. They actually admit to be interpreted as matrices (infinite, but 
with discrete entries), if one rewrites them in the form $F(b, b')$ where 
$b \sim b'$ (actually $b'= b+an$). For such functions we are about to define 
a multiplication structure, which, of course, will be inspired by the usual 
matrix product: 
\begin{eqnarray}\label{c8} \displaystyle{
H \equiv F * G 
} \end{eqnarray}
\begin{eqnarray} \label{c9} \displaystyle{
H(b, m) = \sum_{n} {F(b, n) \:  G(b+an, m-n)} 
} \end{eqnarray}
\par The matrices of (\ref{c9}) are actually parametrized by $b$, but if we 
replace $b$ with $b'= b+ ap$, this results, at the level of the matrix 
product, in a relabeling (by positions) of rows and columns of an infinite 
matrix. In other words, they are actually dependent on the leaf only. 
\par It is straightforward to check the matching of the $*$ definition in (\ref{c8}-\ref{c9})and 
of the matrix product in (\ref{1.8}) and, thus, the equivalence of the two descriptions.

\subsection{D-branes and non-commutative Yang-Mills theory}
The main purpose of the present section is to connect the parameters of a super 
Yang-Mills theory on a fuzzy torus \cite{CR} (in particular the value of $N$) 
with the 
geometrical viewpoint of the non-commutative foliation. The first step 
will be to just assume $\theta = 1/N$. We will later handle the more general case 
of rational non-commutative parameter:
\begin{eqnarray} \label{cqq} \displaystyle{ 
\theta \equiv \frac{p}{N} \: \: \: \: \: \: \: \: \: \: \:  
p, \: N \: {\mathrm {relatively \: prime}}
} \end{eqnarray} 

\begin{figure}
\centerline{\psfig{file=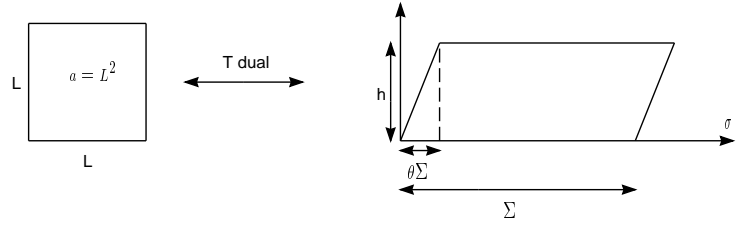,width=10cm}}
\vspace*{8pt}
\caption{T-duality.}
\end{figure}

Following the approach of \cite{CDS,DH}, we consider a weakly coupled IIA 
theory\footnote{We refer to Seiberg's \cite{S}
view of matrix theory \cite{BFSS,CH}, that is, we replace the lightlike 
compactification with a compactification along a spacelike circle of shrinking 
radius.}  compactified on a square torus, with $\tau=i$ and area 
$a=\Sigma^{-1} h$. 
We take a non zero value $\theta$ for the background 2-form potential. 
It will be convenient (for T 
duality purposes) to choose as parameters of the system 
\begin{itemize}
\item{$\tau \equiv i$}
\item{$P := \theta + i\Sigma^{-1} h$}
\end{itemize}
\par Let's do now a T-duality transformation along one cycle of the torus, 
let's say, the ``1'' 1-cycle. Such a transformation interchanges $\tau$ and 
$P$:
\begin{eqnarray} \displaystyle{ 
\begin{array}{ccc}
{ \begin{array}{l}
\left \{
\begin{array}{l}
\tau = i  \\
P = \theta + i \Sigma^{-1} h 
\end{array}
\right .  
\end{array}
} & 
\stackrel{ {\mathrm {T \: duality } } }{ \longleftrightarrow }
& { \begin{array}{l}
\left \{
\begin{array}{l}
\tau' = \theta + i \Sigma^{-1} h \\
P' =  i  
\end{array}
\right . 
\end{array} }
\end{array}
} \end{eqnarray} 
Let the axis, corresponding to the ``1'' cycle along which T-duality is performed, be called 
$\sigma$, with $0 \le \sigma \le \Sigma$. We summarize the situation in Fig.~2.

\par What happens to a D0-brane under this duality? We will obtain a D1-brane 
oriented along the ``1'' direction of the new torus. (See Fig.~3.) Now 
suppose we have a fundamental string whose ends are on the D1-brane but which 
is wound $w$ times along the ``2'' 1-cycle. If we try to impose a constraint 
of minimal length (provided the winding number $w$ is fixed) and if we imagine 
to ``open up'' the torus and refer to a tiling of the plane made by its 
copies (see Fig.~4), we realize that the two points where the fundamental 
string is attached to the D-string are separated by a distance $w \Sigma\theta$ 
along a straight line almost perpendicular to the D-string: 
they are equivalent points in the sense of the foliated torus construction 
described before. 

\begin{figure}
\centerline{\psfig{file=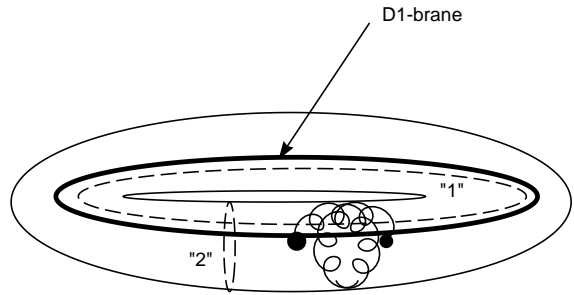,width=8cm}}
\vspace*{8pt}
\caption{D1 Brane along the ``1'' direction.}
\end{figure}

\begin{figure}
\centerline{\psfig{file=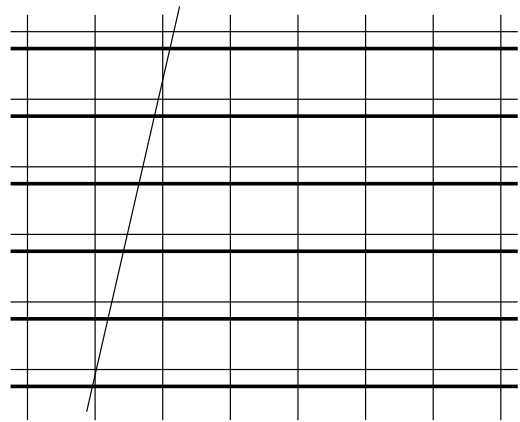,width=7cm}}
\vspace*{8pt}
\caption{``Opened up'' Torus.}
\end{figure}

\par Consider, at this stage, the (nonlocal) field operators which create and 
annihilate such strings. They are fields whose arguments are the two points, $\sigma$ and  
$\displaystyle{\sigma + w \frac{\Sigma}{N}}$ ($w \in {\mathbb Z}$), of the D-string 
connected by the ``minimal length'' fundamental strings. As these two points
are related by the equivalence relation of the 
foliated torus, it is natural to interpret such fields as 
objects which live in the non-commutative algebra of the foliated torus.:
\begin{eqnarray} \label{-1} \displaystyle{ 
\phi(\sigma, w) 
} \end{eqnarray} 
To symmetrize the aspect of the above field, one might Fourier transform 
with respect to the periodic coordinate $\sigma$, thus obtaining a function 
of two integers: 
\begin{eqnarray} \label{0} \displaystyle{ 
\tilde{\phi} (k, w) \: \: \: \: \: \: \: \: \: \: \: k, \, w \in {\mathbb Z}
} \end{eqnarray} 
The energy of such a mode is 
\begin{eqnarray} \label{1} \displaystyle{ 
E(k, w) = \left( \frac{1}{\Sigma^2} \left[ k^2 + w^2 \right] 
\right)^{\frac{1}{2}} 
} \end{eqnarray} 
Let us now imagine of ``folding'' the circle according to the equivalence 
relation induced by the foliation (see Fig.~5). 
The small circle has now length $\displaystyle{\frac{\Sigma}{N}}$ and its 
periodic coordinate $x$ satisfies $\displaystyle{0 \le x \le \frac{\Sigma}{N} 
}$\,. 

\begin{figure}
\centerline{\psfig{file=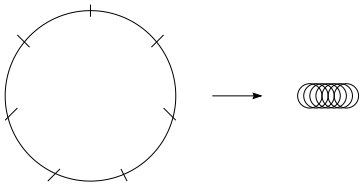,width=5cm}}
\vspace*{8pt}
\caption{Circle Folding.}
\end{figure}

\par We should also notice that the choice of $p$ in the numerator of $\theta$ in (\ref{cqq}) only 
corresponds to a relabeling of the $N$ sectors, which are rearranged in a 
permuted order. (See Fig.~6). 
This is a consequence of the (trivial) fact that, given $p$, $N$ 
relatively prime, no one of the numbers 
\begin{eqnarray} \nonumber \displaystyle{ 
p, \: 2p, \:, 3p, \: ... \: (N-1)p 
} \end{eqnarray} 
is a multiple of $N$ and thus 
\begin{eqnarray} \nonumber \displaystyle{ 
\{ 0, \: (p)_{{\mathrm{mod}}\: N}, \: (2p)_{{\mathrm{mod}}\: N} \:, ... \} 
} \end{eqnarray} 
is a set of $N$ integer numbers belonging to $[0, N]$ and all different. 
Thus choosing $p/N$ instead of $1/N$ results only in a relabeling of the 
equivalence relation: the ``arches'' are superimposed in a different order. 

\begin{figure}
\centerline{\psfig{file=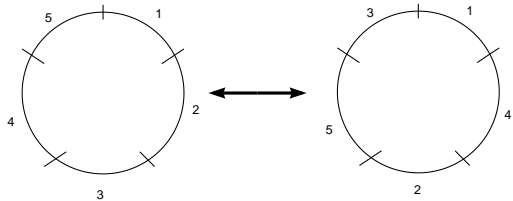,width=7cm}}
\vspace*{8pt}
\caption{Permutation.}
\end{figure}

\par We are now going to split the winding and the momentum modes into a 
part which is an integer multiple of $N$ and the remainder (see Fig.~7). 
First we define an appropriate splitting of the integer $k$ of 
equation (\ref{0}): 
\begin{eqnarray} \label{*bis} \displaystyle{ 
k= KN + q \: \: \: \: \: \: \: \: \: \: \: \: \: \: \: \: K, q \in {\mathbb Z } 
\: \: \: \: \: \: \: \: 0 \le q < N
} \end{eqnarray} 
\par We simultaneously replace the fields $\phi$ of eq.~(\ref{-1}) with 
matrix valued fields 
\begin{eqnarray} \label{A} \displaystyle{ 
\phi_{ab} (x, W) \: \: \: \: \: \: \: \: \: \: \: a, b = 1, ...N 
} \end{eqnarray} 
where the index $a$ (resp.~$b$) tells us in which of the $N$ intervals 
(of length $\displaystyle{\frac{\Sigma}{N}}$ each) the string begins 
(resp.~ends) and the integer $W$ is the winding number along the big circle 
of length $\Sigma$. 

\begin{figure}
\centerline{\psfig{file=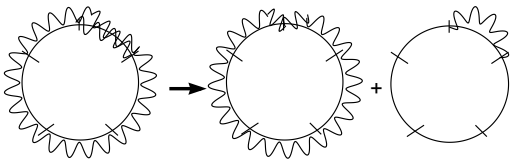,width=7cm}}
\vspace*{8pt}
\caption{Winding and Momentum Modes.}
\end{figure}

\par The relation with $w$ is given by 
\begin{eqnarray} \displaystyle{ 
w= NW + (a-b) 
} \end{eqnarray} 
We can now rewrite the energy (\ref{1}) in terms of the capital variables: 
\begin{eqnarray} \label{spectrum} \displaystyle{ 
E^2 = \frac{1}{\Sigma^2} \left( (NK+q)^2 + (NW+(a-b))^2  \right) 
} \end{eqnarray} 
\par We now wish to reproduce the energy spectrum (\ref{spectrum}) with a 
description on a small torus of size $\Sigma / N$. On such torus we will 
have two directions, $x$ and $y$ respectively. The description (\ref{A}) of 
the matrix fields can yield a description in one more direction if, as 
usual, we wish to interpret the winding mode as a Kaluza-Klein momentum in 
an additional direction; that is, Fourier transformation with respect to 
$W$ gives 
\begin{eqnarray} \displaystyle{ 
\phi_{ab} (x, y) 
} \end{eqnarray} 
\par In order to reproduce the spectrum (\ref{spectrum}) we will introduce a 
background $U(N)$ Yang-Mills field. We will also have to introduce non-trivial boundary 
conditions for the fields when trasporting them around 
the $x$ axis. The first couple of condition we impose are 
\begin{eqnarray} \displaystyle{ 
\phi_{a, b} (x + \frac{\Sigma}{N}, y)  = \phi_{a+1, b+1} (x, y)
} \end{eqnarray} 
\begin{eqnarray} \displaystyle{ 
\phi_{ab} (x, y+ \frac{\Sigma}{N}) = \phi_{ab} (x, y)
} \end{eqnarray} 
that is, if we move along $x$ direction making a complete turn on the small 
circle we get a unit shift of both indexes (of the ``begin'' and ``end''  
sectors on the big circle), while the $y$ direction is associated to the 
``big'' winding number. In matrix language  
\begin{eqnarray} \displaystyle{ 
\phi(x + \frac{\Sigma}{N}, y)  = U^\dagger \phi(x, y) U 
} \end{eqnarray} 
\begin{eqnarray} \displaystyle{ 
\phi(x, y+ \frac{\Sigma}{N}) = \phi(x, y)
} \end{eqnarray} 
where $U$ is the $N \times N$ shift matrix given in (\ref{shift}).
Along the $y$ direction we wish to introduce Wilson loops, in order to mimic 
the fractional contribution to the momentum which we encountered in 
eq.~(\ref{spectrum}): 
\begin{eqnarray} \displaystyle{ 
W(x) = exp \left( i \oint A_y (x,y) \: dy  \right)
} \end{eqnarray} 
We assume $A_x =0$ and $A_y$ independent of $y$: 
\begin{eqnarray} \label{c1} \displaystyle{ 
W(x) = exp \left( i \frac{\Sigma}{N} A_y(x)  \right) 
} \end{eqnarray} 
We put in a background vector potential: 
\begin{eqnarray} \label{c2} \displaystyle{ 
W (x) = exp \left( i \frac{x}{\Sigma}   \right) V 
} \end{eqnarray} 
where the matrix $V$ is as in (\ref{phases}).
Notice that the background satisfies the same conditions of $\phi$ if we  
make a complete turn on the small circle: 
\begin{eqnarray} \displaystyle{ 
\Gamma(x + \frac{\Sigma}{N}) = U^{\dagger} \Gamma (x) U 
} \end{eqnarray} 
\par From (\ref{c1}) and (\ref{c2}) it follows immediately for the vector 
potential 
\begin{eqnarray} \displaystyle{ 
A_y = \frac{x}{\Sigma^2} N \: I + \frac{1}{\Sigma} 
\left( 
\begin{array}{cccccc}
\ddots & 0 & 0 & 0 & 0 & 0 \\
0 & -2 & 0 & 0 & 0 & 0 \\
0 & 0 & -1 & 0 & 0 & 0 \\
0 & 0 & 0 & 0 & 0 & 0 \\
0 & 0 & 0 & 0 & 1 & 0 \\
0 & 0 & 0 & 0 & 0 & \ddots \\
\end{array}
\right) 
} \end{eqnarray} 
where $I$ is the $N \times N$ identity matrix. 
\par The vector potential can thus be split in an ``abelian'' and in a  
``non abelian'' part. One should notice how the form of the abelian gauge 
field corresponds to a unit of abelian magnetic flux through the torus. 
\par Now let's consider the free part of the action for 
the scalars 
\begin{eqnarray} \displaystyle{ 
{\cal L} = \int \: dx \: dy \: Tr (\dot{\phi}^2 - (\nabla \phi)^2 )
} \end{eqnarray} 
where the covariant derivative is defined as usual 
\begin{eqnarray} \displaystyle{ 
\nabla \phi = \partial \phi + i [A, \phi] 
} \end{eqnarray} 
Let us consider separately the spectra of $\nabla_x$ and $\nabla_y$. 
\par First of all, the spectrum of $\nabla_x \equiv \partial_x$ has to be 
evaluated on a circle of length $\displaystyle{\frac{\Sigma}{N}}$. But 
actually we imposed a condition which allows $\phi$ to return to its value 
only after $N$ turns, so the system behaves as the circle was effectively 
of size $\Sigma$. The spectrum will be then 
\begin{eqnarray} \displaystyle{ 
\frac{i}{\Sigma} [KN + p] \: \: \: \: \: \: \: \: \: \: \: p \in {\mathbb{Z}} 
\, , \: \: 0 \le p \le N-1  
} \end{eqnarray} 
with the same decomposition of eq.~(\ref{*bis}).   
\par Second, let's turn our attention to $\displaystyle{\nabla_y =  
\partial_y + i [A_y, \cdot ]}$\,. The derivative piece has the usual 
spectrum on the circle 
\begin{eqnarray} \displaystyle{ 
i \frac{NW}{\Sigma} \: \: \: \: \: \: \: \: \: \: \: W \in {\mathbb{Z}}
} \end{eqnarray} 
since $\displaystyle{\phi(y + \frac{N}{\Sigma})= \phi(y)}$. The commutator  
(for which the ``abelian'' part of the vector potential can be dropped) is 
rewritten 
\begin{eqnarray} \displaystyle{ 
[A_y, \phi]_{a,b} = \frac{1}{\Sigma} (a-b) \phi_{a,b}
} \end{eqnarray} 
and the spectrum of $\nabla_y$ is 
\begin{eqnarray} \displaystyle{ 
\frac{i}{\Sigma} \left[ NW + (a-b) \right] 
} \end{eqnarray} 
Since the equation of motion of $\phi$ is 
\begin{eqnarray} \displaystyle{ 
\ddot{\phi}  = ( \nabla_x^2 + \nabla_y^2 ) \phi 
} \end{eqnarray} 
the spectrum will be 
\begin{eqnarray} \displaystyle{ 
E^2 = \left[  \left( \frac{KN + p}{\Sigma} \right)^2 + \left( 
\frac{WN + (a-b)}{\Sigma} \right)^2  \right] 
} \end{eqnarray} 
in agreement with eq.~(\ref{spectrum}).

\subsection{Interaction terms}
We want to proceed to an 
explicit exhibition of how the star product works as a substitute of matrix 
multiplication and to discuss the role of the numerator of the fraction 
$\theta \equiv p/N$. To isolate our main point, we will refer to the quartic scalar interaction 
term, even if the whole procedure extends straightforwardly to the case of generic interactions.
\par In the 2+1 dimensional Connes-Douglas-Schwartz theory over the 
``big'' torus, the form of the quartic terms is 
\begin{eqnarray} \label{quartic} \displaystyle{ 
\int_0^\Sigma {dX \: dY \: [\phi^i * \phi^j - \phi^j * \phi^i  ]^2 }
}\end{eqnarray} 
where $X$, $Y$ are coordinates on the ``large'' torus and the star product 
is defined as  
\begin{eqnarray} \displaystyle{ 
F * G = F(x,y) \: e^{i \frac{\theta}{2} (    
{ \stackrel{\leftarrow}{\partial}}_x \vec{\partial}_y 
- { \stackrel{\leftarrow}{\partial}}_y \vec{\partial}_x  )} \: G(x,y) 
} \end{eqnarray} 
For the moment we will assume $\theta \equiv 1/N$. 
\par To evaluate this, we go to the Douglas-Hull representation by Fourier 
transforming with respect to $y$: 
\begin{eqnarray} 
F * G &=& \sum_{n, m} F(x, n) \: e^{i n y} \: e^{i \frac{\theta}{2} (    
{ \stackrel{\leftarrow}{\partial}}_x \vec{\partial}_y        
- { \stackrel{\leftarrow}{\partial}}_y \vec{\partial}_x  )}  
\: \tilde{G}(x, m) \: e^{imy}\nonumber\\
&=& \sum_{n, m} {F(x,n) \: e^{-m \theta /2 \vec{\partial}_x }   \: 
e^{n \theta / 2 \vec{\partial}_x }  \: G(x,n) \: e^{i(n+m)y } } = \nonumber\\
&=& \sum F(x- \frac{m \theta}{2}, n) \: G(x+ \frac{n \theta}{2}, m) 
\: e^{i(n+m)y} 
\end{eqnarray} 
\par We now recall the intuitive picture of $F(x,n)$ as a ``string''
attached at $x$ and with length $n \theta$ along the $x$ axis (see Fig.~8). 
If we regard the strings whose 
endpoints belong to the same leaf (i.~e.~are equivalent with respect to the 
leaf induced relation of equivalence) as matrices (since the endpoints 
of $F$ and $G$ coincide), then the Fourier transform of $F*G$ is 
exactly the matrix product: 
\begin{eqnarray} \displaystyle{ 
\sum_n F(x- \frac{k-n}{2} \theta, n) \: G(x + \frac{n \theta}{2}, (k-n))   
} \end{eqnarray} 
(remember $m=k-n$). 

\begin{figure}
\centerline{\psfig{file=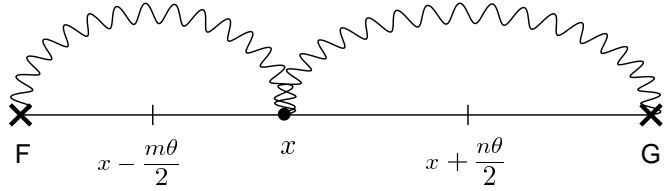,width=9cm}}
\vspace*{8pt}
\caption{$F(x,n)$ as a ``string''.}
\end{figure}

\par Notice that a given leaf corresponds to a point on a small circle. 
It was actually to the purpose of removing (this particular kind of) 
non locality that the small circle has been introduced. 
\par Similarly, expressions like 
\begin{eqnarray} \displaystyle{ 
\int (F*G)(H*K) \: dx \: dy 
} \end{eqnarray} 
translate in matrix language 
\begin{eqnarray} \displaystyle{ 
\int_0^{\frac{\Sigma}{N}} Tr \: (FG)(HK)  
} \end{eqnarray} 
Thus, going back to the action term (\ref{quartic}) we obtain  
\begin{eqnarray} \displaystyle{ 
Tr \: \int_0^{\frac{\Sigma}{N}} [\phi^i, \phi^j]^2 \: dx \: dy   
} \end{eqnarray} 
that is, our usual Yang-Mills terms. 
\par Let us now discuss what happens if $\theta = p/N$, $p$ and $N$ 
relatively prime. We have already discussed how this amounts to a 
``relabeling'' of the sectors of the circle by means of a permutation of 
their indices. However, it is important to point out what happens to the 
``boundary condition'' $U$ and to the ``Wilson loop'' $V$ matrices. 
\par The first point to realize is that, whatever the permutation of 
sectors may be, the ``periodic'' boundary conditions will not change. This 
is because, no matter what happens to the interactions, the free 
evolution will move along the interval in the original order; the free part 
of the lagrangian on the big circle is certainly local. 
\par What happens to the Wilson loop? If we permute the intervals 
\begin{eqnarray} \displaystyle{ 
\begin{array}{c} 
\begin{array}{lll}
1 & \longrightarrow & 1 \\
2 & \longrightarrow & (1+p)_{{\mathrm{mod}\:}N} \\ 
3 & \longrightarrow & (1+2p)_{{\mathrm{mod}\:}N} \\    
4 & \longrightarrow & (1+3p)_{{\mathrm{mod}\:}N} \\    
\end{array}        \\ 
\ldots \ldots \ldots 
\end{array} 
} \end{eqnarray} 
we will replace eq.~(\ref{c2}) with 
\begin{eqnarray} \displaystyle 
W(x)= exp \left(  i p \frac{x}{\Sigma} \right)  V^p 
\end{eqnarray} 
The non abelian part of the Wilson loop $V$ is replaced with 
\begin{eqnarray} \displaystyle{ 
V^p = \left( 
\begin{array}{cccccc}
1 & 0 & 0 & 0 & 0 & 0 \\
0 & \xi^{(1+p)_{{\mathrm{mod}\:}N} -1}  & 0 & 0 & 0 & 0 \\
0 & 0 & \xi^{(1+2p)_{{\mathrm{mod}\:}N} -1} & 0 & 0 & 0 \\
0 & 0 & 0 & \xi^{(1+3p)_{{\mathrm{mod}\:}N} -1} & 0 & 0 \\
0 & 0 & 0 & 0 & \ddots & 0 \\
0 & 0 & 0 & 0 & 0 & \ddots \\ 
\end{array}
\right) 
\: \: \: \: \: \: \: \xi= e^{2 \pi i /N}
} \end{eqnarray} 
The abelian vector potential is now $p$ times larger, and so is the flux. 
\par Also the commutation relation is accordingly modified:  
\begin{eqnarray} \displaystyle{ 
UW= WU \, \xi^p 
} \end{eqnarray} 
(the rewriting in terms of the Wilson loop is allowed since the abelian 
part gives no contribution). 
\par To summarize, the relations in the general case are 
\begin{eqnarray} \label{annidata} \displaystyle{ 
\begin{array}{c}
\theta= \frac{p}{N} \\ \\  
\left\{ \begin{array}{l} 
\phi(x+ \frac{\Sigma}{N}, y) = U^\dagger \phi(x,y) U \\ 
\phi(x, y+ \frac{\Sigma}{N}) = \phi(y) 
\end{array} \right. \\ \\ 
W= V^p \\  \\ 
UW= WU \, \xi^p 
\end{array}
} \end{eqnarray} 
Moreover, the abelian part of the vector potential carries now $p$ units 
of magnetic flux. 

\subsection{Coupling constant rescaling}
\par It might be interesting to derive the behaviour of the coupling 
constant $g_{YM}$ along the process of ``folding'' the big circle into 
the small one. We know (cfr.~for example \cite{review}) that the 
Yang-Mills coupling for a 2+1 dim gauge theory describing compactification on 
a 2-torus is
\begin{eqnarray} \displaystyle{ 
g_{YM}^2 = \frac{1}{\Sigma} \frac{l_{11}^3}{L^3} 
} \end{eqnarray} 
This is the coupling constant of the non commutative geometry on the 
``big'' torus. The transition to the small torus yields a factor of $1/N$: 
\begin{eqnarray} \displaystyle{ 
\tilde{g}_{YM}^2 = \frac{1}{N \Sigma} \frac{l_{11}^3}{L^3}
} \end{eqnarray} 
Thus $g^2 N$ is kept fixed during the passage to the ``small'' torus. 

\subsection{Summary}
To summarize, we have now a prescription for the link between a super Yang 
Mills theory with large N and a non-commutative geometry with $\theta$ 
``almost irrational''. \\

\begin{itemize}
\item{Take $\theta$ and approximate it with an irreducible fraction  
$\displaystyle{\frac{p}{N}}$}
\end{itemize}
\begin{eqnarray} \nonumber \displaystyle{ 
\Updownarrow 
} \end{eqnarray} 
\begin{itemize}
\item{Build a $U(N)$ gauge theory on a torus of size 
$\displaystyle{\frac{\Sigma}{N}}$ and choose as boundary conditions those   
which are explicit in eq.~(\ref{annidata}) with $p$ units of magnetic flux.}
\end{itemize}
\vspace{0.2cm} 
Thus we find that abelian gauge theory on a non-commutative torus is equivalent 
to an appropriate limit of non abelian gauge theory on a rescaled commutative 
torus. 

\section{Wilson Lines on the Fuzzy Torus}
In this section we will be interested in the construction of gauge
invariant Wilson Loops in a regularized version of \nc \ gauge
theory \cite{FT}. The theory we are using and much of our results  have 
been discussed for the first time by
Ambjorn,  Makeenko,  Nishimura and
Szabo \cite{AMNS}. Here we give much less general, but hopefully more simple,
presentation of those results.

The regularized theory is a \nc \ version of lattice gauge
theory on the Fuzzy Torus. It is patterned after the
Hamiltonian form of lattice gauge theory  \cite{KS}.

The lattice version is an especially
intuitive formulation of the non-perturbative theory.
For illustrative purposes we will concentrate on the
Abelian theory in $2+1$ dimensions. The generalization to higher
dimensions and non-abelian gauge groups is straightforward.
Our main focus will be on defining the gauge
invariant quantities of the theory including closed and open
Wilson lines and in formulating the theory of matter in the
fundamental representation of the non-commutative algebra of
functions.

\subsection{Lattice Gauge theory on the Fuzzy Torus}
The \ft  \ is analogous to a periodic lattice. If we introduce coordinates
\be
y = q R\,,\qquad x = p R
\label{xy}
\ee
such that
\be
[y,x]=i\theta R^2
\ee
and moreover choose $\theta=2\pi/N$, the lattice spacing is
\be
 a=2 \pi R/N.
\label{1.9}
\ee
This is because the Fourier expansion
in eq.(\ref{1.7}) has only a finite number of terms. In other words
there is a largest momentum in each direction
\be
p_{max} = 2\pi(N-1)/R
\ee
Thus the \ft \ has both an infrared cutoff length $R$ and an
ultraviolet cutoff length $2 \pi
R/N$

The operators $U,V$ function as shifts on the periodic lattice.
Using the last of eq's(\ref{1.1}) one easily finds
\bea
U\phi(U,V)\udag&=&\phi(U, Ve^{i \theta})\cr
\udag \phi(U,V)U&=&\phi(U, Ve^{-i \theta})\cr
V\phi(U,V)\vdag&=&\phi(Ue^{-i \theta}, V)\cr
\vdag \phi(U,V)V&=&\phi(Ue^{i \theta}, V)\cr
\eea
More generally
\be
U^n V^m \phi(U,V)
{\vdag}^m {\udag}^n =
\phi(U e^{-im\theta}, V e^{in\theta} )
\ee

The rule for integration on the \ft \ is simple.
\be
\int U_{mn} = 4 {\pi}^2 R^2 \delta_{m0} \delta_{n0}
\ee
Noting that
\be
Tr U_{mn} =N\delta_{m0} \delta_{n0}
\ee
we make the identification
\be
\int F(U,V) = \frac{4 {\pi}^2 R^2}{N} Tr F(U,V)
\ee

\par In what follows  we will work in the temporal gauge in which the
time component of the vector potential is zero.

Let us introduce gauge fields on the fuzzy torus in analogy with
the link variables of lattice gauge theory \cite{DH}. We will explicitly
work with the gauge group $U(1)$. The link variable in the $x,y$
direction is called $X,Y$. The link variables are unitary
\bea
\xdag X&=&1 \cr
\ydag Y&=&1
\eea
The gauge invariance of the theory is patterned on that of lattice
gauge theory. Let $Z$ be a unitary, time independent  function of $U,V$,
$\zdag
Z=1$. The gauge transformation induced by $Z$ is defined to be
\bea
X'&=&Z(U,V)X(U,V) \zdag (U\ep , V) \cr
Y'&=&Z(U,V)Y(U,V) \zdag (U , V\ep)
\eea
or
\bea
X'&=&Z X \ \vdag \zdag V \cr
Y'&=&ZY U \zdag \udag
\label{2.3}
\eea

Let us now construct Wilson loops by analogy with the conventional
lattice construction. We will give some examples first. A Wilson
line which winds around the x-cycle of the torus at a fixed value
of $y$ is given by
\bea
W_x &=& Tr X(U,V)X(U\ep,V)X(U\epp,V)..X(U\epn,V)\cr
    &=& Tr (X\vdag)^N
\eea
Similarly
\be
W_y=Tr (YU)^N
\ee
These expressions are gauge invariant under the transformation in
eq.(\ref{2.3}).

Another example of a Wilson loop is the analogue of the plaquette
in lattice gauge theory. It is given by
\bea
{\cal{P}} \q Tr X(U,V)Y(U\ep ,V)\xdag(U,V\ep)\ydag(U,V)\cr
\q Tr (X)(\vdag Y V)(U\xdag \udag)( \ydag)\cr
\q \exm Tr(X\vdag)(YU)(V\xdag)(\udag \ydag)
\label{2.6}
\eea

The general rule involves drawing a closed  oriented chain formed
from directed links. A step in the positive (negative) $x$ direction is
described by the link operator $X\vdag$ ($V \xdag$). Similarly a step in
the positive (negative) $y$ direction gives a factor $YU$ ($\udag
\ydag$). The link operators are multiplied in the order specified
by the chain and the trace is taken. In addition there is a factor
$e^{-i A
\theta}$ where $A$ is the signed Area of the loop in units of the
lattice spacing. For a simple  contractable clockwise oriented loop with
no crossings, $A$ is just the number of enclosed plaquettes.

A simple Lagrangian for the gauge theory  can be formed from
plaquette operators and kinetic term involving time derivatives. The
expression
\be
Tr \dot {\xdag} \dot {X}+\dot {\ydag} \dot {Y}
\ee
is quadratic in time derivatives and is gauge invariant.
Again, following the model of lattice gauge theory \cite{KS} we choose the
action
\be
{\cal{L}} =
\frac{4\pi^2 R^2}{g^2 a^2 N}
Tr\left [ \dot {\xdag} \dot {X}+\dot {\ydag} \dot {Y}
+\frac{\exm}{a^2}  (X\vdag)(YU)(V\xdag)(\udag \ydag)+cc \right]
\ee

Evidently the operators $X \vdag$ and $YU$ play an important
role.  We therefore define
\bea
\x \q X\vdag \cr
\y \q YU
\eea
These operators transform simply under gauge transformations:
\bea
\x &\to& Z\x Z^{\dag} \cr
\y  &\to& Z\y Z^{\dag}
\eea

The action is now written in the form
\be
{\cal{L}}=\frac{4\pi^2 R^2}{g^2 a^2 N}
Tr\left [ \dot {\xd} \dot {\x}+\dot {\yd} \dot {\y}
+\frac{\exm}{a^2} {\x}{\y}{\xd} {\yd}+cc \right]
\ee
or using eq.(\ref{1.9})
\be
{\cal{L}}=\frac{N}{g^2}
Tr\left [ \dot {\xd} \dot {\x}+\dot {\yd} \dot {\y}
+\frac{\exm}{a^2} {\x}{\y}{\xd} {\yd}+cc \right]
\label{2.12}
\ee
In this form the action is equivalent to that of a $U(N)$ lattice
gauge theory formulated on a single plaquette but with periodic
boundary conditions of a torus. This appears to be a form of
Morita equivalence \cite{dani}.

If the coupling constant is small, the ground state is determined
by minimizing the plaquette term in the Hamiltonian. This is done
by setting
\be
\exm {\x}{\y}{\xd} {\yd}=1
\ee
Up to a gauge transformation the unique solution of this equation
is
\bea
\x &=&\vdag  \cr
\y \q U
\label{2.14}
\eea
or
\be
X=Y=1
\ee

\subsection{Open Wilson Loops}

Thus far we have constructed closed Wilson loops. Recall that  the
construction involves taking a trace. This is the analogue of
integrating the location of the Wilson loop over all space. In
other words the closed Wilson Loop carries no spatial momentum. In
a very interesting paper Ishibashi,  Iso,  Kawai and Kitazawa
\cite{IIKK} have argued that there
exist gauge invariant operators which correspond to specific
Fourier modes of open Wilson lines. These objects are very closely
related to the growing dipoles of \nc \ field theory whose size
depends on their momentum \cite{SJ,SJ2,BS}. Das and Rey \cite{DR} have shown
that these
operators are a complete set of gauge invariant operators. Their
importance has been further clarified by  Gross,  Hashimoto and Itzhaki
\cite{GHI}.

Let us consider the simplest example of an open Wilson line, ie, a
single  link variable, say $X$. From eq.(\ref{2.3}) we see that $X$ is
not gauge invariant.  But now consider $X \vdag =\x$. Under gauge
transformations
\be
\x  \to Z\x Z^{\dag}
\ee
Evidently the quantity
\be
Tr X \vdag =Tr \x
\ee
is gauge invariant. Now using (\ref{commrel}) and (\ref{xy}) we identify this quantity as
\be
Tr X \vdag =\frac{N}{4 \pi^2 R^2}\int X e^{\frac{-iy}{R}} d^2 x
\ee
Thus we see that a particular Fourier mode of $X$ is gauge
invariant.

Let us consider another example in which an open Wilson line
consist of two adjacent links, one along the $x$ axis and one
along the $y$ axis.
\be
X \vdag YV =\x \y \udag V
\ee
Multiplying by $\vdag U$ and taking the trace gives
\be
Tr(X \vdag YV)\vdag U = Tr \x \y = gauge \ invariant
\ee
But we can also write this as
\be
\frac{N}{4 \pi^2 R^2}\int d^2x (X \vdag YV)e^{\frac{-iy}{R}}e^{\frac{-ix}{R}}
\ee
In other words it is again a Fourier mode of the open Wilson line.
In general the particular Fourier mode is related to the
separation between the endpoints of the \wl by the same relation
as that in \cite{SJ2,BS} where it was shown that a particle in \nc \ field
theory is a dipole oriented perpendicular to it momentum with a
size proportional to the momentum.

\subsection{Fields in the Fundamental Representation}

In this section we will define fields in the fundamental
representation of the gauge group. For simplicity we consider
non-relativistic
particles. Let us begin with what we do
$not$ mean by particles in the fundamental. Define a complex valued
field $\phi$ that takes values in the $N \times N$ dimensional matrix
algebra
generated by $U,V$. The gauge transformation properties of $\phi$ are
given by
\be
\phi  \to Z \phi.
\ee
Note that this is left multiplication by $Z$ and not conjugation.
The field $\phi$ carries a single unit of abelian gauge charge. Although
the field
has two indices in the $N$ dimensional space the gauge transformations
only act on the
left index.

An obvious choice of gauge invariant
"hopping" Hamiltonian would be
\be
H \sim   Tr \phi \da X \vdag \phi V  \phi \da X U \phi \udag +cc
\label{4.2}
\ee
In a non-abelian theory a similar construction can be carried out
for quark fields in the fundamental.

We shall mean something different  by fields in the fundamental.
Such fields have only $one$ index. They are vectors rather than matrices
in the Hilbert space that the represents
the algebra of functions. In the present case they are $N$
component complex vectors $|\psi\rangle$. These fields represent
particles moving in a strong magnetic field which are frozen into
the lowest Landau level.

Consider the case of non-relativistic particles moving on
the \nc \ lattice. The conventional lattice action would be
\be
L=L_0 - L_h
\ee
where
\be
L_0 = i\left(\langle\dot \psi \da |\psi \rangle  -cc     \right)
\label{4.4}
\ee
and  $L_h$ is a hopping Hamiltonian. The natural \nc \ version of
the hopping term is
\be
L_h =\frac{1}{a} \langle \psi| X \vdag +YU -2 |\psi\rangle + cc
\ee
The presence of the link variables $X,Y$ is familiar from ordinary
lattice field theory and the $\vdag , U$ are  the shifts which
move $\psi$. We may also write the hopping term as
\be
L_h = \frac{1}{a} \langle \psi| \x +\y | -2 \psi\rangle + cc
\label{4.6}
\ee

Combining (\ref{2.12}), (\ref{4.2}) and (\ref{4.4})

\be
{\cal{L}}=\frac{N}{g^2}
Tr\left [ \dot {\xd} \dot {\x}+\dot {\yd} \dot {\y}
+\frac{\exm}{a^2} {\x}{\y}{\xd} {\yd}+cc \right]
+ i\langle\dot \psi \da |\psi \rangle
+ \frac{1}{a}\langle \psi| \x +\y -2 |\psi\rangle + cc
\ee
Let us consider hopping terms in (\ref{4.6}). In the limit of weak coupling
we may use eq.(\ref{2.14})
to give
\be
L_h= \frac{1}{a}\langle \psi| \vdag +U -2 |\psi\rangle + cc
\ee
To get some idea of the meaning of this term let us use eqs.(\ref{commrel})-(\ref{xy})
and expand the exponentials.
\bea
L_h &=& \frac{1}{a} \langle \psi| \frac{(x^2 +y^2)}{R^2}|\psi\rangle + cc = \nonumber\\
 &=& \frac{1}{a} \langle \psi| (p^2+q^2 )\theta|\psi\rangle + cc
\eea
As $[p,q]=i\theta$, we recognize this term as a
harmonic oscillator hamiltonian with an in spectrum of levels spaced by $\theta \sim N^{-1}$. 
Evidently, in this approximation the particles move in quantized circular orbits around the origin.

This phenomena is related to the fact that the fundamental
particles behave like charged particles in a strong magnetic field
and are frozen into their lowest Landau levels. Furthermore the
LLL's are split by a force attracting the particles to $x=y=0$.
This has a natural interpretation in matrix theory in which the
same system appears as a 2-brane and 0-brane with strings
connecting them \cite{AB}.

\subsection{Rational Theta}

Thus far we have worked with eq.(\ref{1.1}) with $\theta =2 \pi /N$. Let
us generalize the construction to the case $\theta = 2 \pi p/N$
with $p$ relatively prime to $N$. We continue to define the fuzzy
torus by eq.(\ref{1.1}). Let us define two  matrices $u,v$ satisfying
\bea
u^{\dag}u&=&v^{\dag}v =1 \cr
u^N&=&v^N =1 \cr
uv &=& vu e^{\frac{2 \pi i \alpha}{N}}
\eea
where
\be
\alpha p =1(mod\N)
\ee
Then it follows that
\bea
U \q u^p \cr
V \q v^p
\eea
satisfies eq.(\ref{1.1}). Furthermore, $u$ and $v^{\dag}$ act as shifts
by distance $2 \pi R/N$;
\bea
u V u^{\dag} \q V \exp\left (\frac{2 \pi i}{N}\right) \cr
v^{\dag} U v \q U \exp\left (\frac{2 \pi i}{N}\right)
\eea
The basic plaquette is now given by
\bea
{\cal{P}} \q Tr(X)(v^{\dag}Y v)(u \xdag u^{\dag})(\ydag) \cr
\q  e^{\frac{2 \pi i \alpha}{N}} Tr(X v^{\dag})(Yu) (v\xdag)(u^{\dag}
\ydag)
\eea

The final expression for action is essentially the same as in
eq.(\ref{2.6}) except that the factor $\ep$ is replaced by
$ e^{\frac{2 \pi i \alpha}{N}}$.

We can now describe one approach to the continuum limit, $a \to 0$. To
get to
such a limit  eq.(\ref{1.9}) requires that $N \to \infty$. We also
want the theta parameter to approach a finite limit. This requires
$p/N$ to approach a limit. for example if we want $p/N \to 1/2$ we
can choose the sequence
($p =n, N=2n+1$) so that $p$ and $N$ remain relatively prime. In
this way $p/N$ can tend to a rational or irrational limit and the
lattice spacing will approach zero.


\section{String Theory and Non-commutative Geometry}
We finally give a simple physical picture of how non-commutative geometry arises
in string theory. As we have discussed in the previous sections, 
the structure of such theories is similar
to that of ordinary gauge theory except that the usual product of
fields is replaced by a ``star product'' defined by
\begin{eqnarray} \displaystyle{
\phi * \chi = \phi(X) \exp\{ -i \theta^{\mu \nu}
\frac{\partial}{\partial X^{\mu}}
\frac{\partial}{\partial Y^{\nu}} \}
\chi(Y)}
\end{eqnarray}
where $\theta^{\mu \nu}$ is an antisymmetric constant tensor. The
effect of such a modification is reflected in the momentum space
vertices of the theory by factors of the form
\begin{eqnarray}
\label{cippa2} \displaystyle{ \exp[i \theta^{\mu \nu} p_\mu q_\nu ]
} \equiv e^{i{p\wedge q}}
\label{2}
\end{eqnarray}
The purpose of this section is to show how these factors arise in an
elementary way. We will begin by describing a simple quantum
mechanical system which is fundamental to our construction. We
then consider string theory in the presence of a D3-brane and a
constant large $B_{\mu \nu}$ field. In the light cone frame the
first quantized string is described by our elementary model. We
use the model to compute the string splitting vertex and show how
the factors in eq.~(\ref{2}) emerge. We then turn to the structure of the
perturbation series for the non-commutative theory in infinite
flat space. We find that  planar diagrams with any number of
loops are identical to their commutative counterparts apart from
trivial external line phase factors. The effects of non-commutativity
in a finite size geometry, as the fuzzy torus of the previous sections, are
more subtle. We comment on this issue towards the end of this section.

\subsection{The model}
\subsubsection{Classical level}
Consider a pair of unit charges of opposite sign in a magnetic
field $B$ in the regime where the Coulomb and the radiation terms
are negligible. 
The coordinates of the charges are $\vec{x_1}$ and
$\vec{x_2}$ or in component form $x_1^i$ and $x_2^i$. The
Lagrangian is
\begin{eqnarray} \displaystyle{
{\cal L} = \frac{m}{2} \left( (\dot{x}_1)^2 + (\dot{x}_2)^2
\right) \:+\: \frac{B}{2} \epsilon_{ij} \left( \dot{x}_1^i x_1^j -
\dot{x}_2^i x_2^j \right) \: -\: \frac{K}{2} (x_1-x_2)^2 }
\end{eqnarray} where the first term is the kinetic energy of the
charges, the second term is their interaction with the magnetic
field and the last term is an harmonic potential between the
charges.
\par In what follows we will be interested in the limit in which
the first term can be ignored. This is typically the case if $B$
is so large that the available energy is insufficient to excite
higher Landau levels \cite{girvin}. 
Thus we will focus on the simplified Lagrangian
\begin{eqnarray} \displaystyle{
{\cal L} = \frac{B}{2} \epsilon_{ij} \left( \dot{x}_1^i x_1^j -
\dot{x}_2^i x_2^j \right) \: -\: \frac{K}{2} (x_1-x_2)^2 }
\end{eqnarray} Let us first discuss the classical system. 
In terms of the center of mass and relative
coordinates $X$, $\Delta$:
\begin{eqnarray} \displaystyle{
\begin{array}{l}
\vec{X} = (\vec{x_1} + \vec{x_2}) /2 \\ \vec{\Delta} = (\vec{x_1}
- \vec{x_2}) /2
\end{array}
} \end{eqnarray} the Lagrangian is
\begin{eqnarray} \label{5} \displaystyle{
{\cal L} = m((\dot{X})^2+(\dot{\Delta})^2) + 2 B \epsilon_{ij}
\dot{X}^i \Delta^j - 2 K (\Delta)^2 } \end{eqnarray} Dropping the
kinetic terms gives
\begin{eqnarray} \displaystyle{
{\cal L} = 2 B \epsilon_{ij} \dot{X}^i \Delta^j - 2 K (\Delta)^2 }
\end{eqnarray} The momentum conjugate to $X$ is
\begin{eqnarray} \label{1.6} \displaystyle{
\frac{\partial {\cal L}}{\partial \dot{X}^i} = 2 B \epsilon_{ij}
\Delta^j = P_i } \end{eqnarray} This is the center of mass
momentum.
\par Finally, the Hamiltonian is
\begin{eqnarray} \label{7} \displaystyle{
{\cal H} = 2 K (\Delta)^2 = 2 K \left( \frac{P}{2B} \right)^2 =
\frac{K}{2B^2} P^2 } \end{eqnarray} This is the hamiltonian of a
nonrelativistic particle with mass
\begin{eqnarray} \displaystyle{
M= \frac{B^2}{K} } \end{eqnarray}

Evidently the composite system of opposite charges
moves like a galileian particle of mass $M$. What is unusual is
that the spatial extension $\Delta$ of the system is related to
its momentum so that the size grows linearly with $P$ according to
eq.~(\ref{1.6}). How does this growth with momentum effect the interactions
of the composite? Let's suppose
charge 1 interacts  locally with an ``impurity'' centered at the origin.
The interaction  has the form
\begin{eqnarray} \label{9} \displaystyle{
V(\vec{x}_1) = \lambda \: \delta (\vec{x}_1) } \end{eqnarray} In
terms of $X$ and $\Delta$ this becomes
\begin{eqnarray} \label{10bis} \displaystyle{
V = \lambda \: \delta (X+\Delta) = \lambda \: \delta(X^i -
\frac{1}{2B} \epsilon^{ij} P_j ) } \end{eqnarray} Note that the
interaction in terms of the center of mass coordinate is nonlocal
in a particular way. The interaction point is shifted by a
momentum dependent amount. This is the origin of the peculiar
momentum dependent phases that appear in interaction vertices on
the non-commutative plane. More generally, if particle 1 sees a
potential $V(x_1)$ the interaction becomes
\begin{eqnarray} \label{1.7bis} \displaystyle{
V\left(X -\frac{\epsilon P}{2B} \right) } \end{eqnarray}

\subsubsection{Quantum level}
The main problem in quantizing the system is to correctly define
expressions like (\ref{1.7bis}) which in general have factor ordering
and other quantum ambiguities. A standard way of solving this problem is 
Weyl ordering: assume that $V$ can be expressed as a Fourier transform
\begin{eqnarray} \label{c14}  \displaystyle{
V(x) = \int dq  \: \tilde{V}(q) e^{iqx} } \end{eqnarray} We can then
formally write
\begin{eqnarray} \displaystyle{
V(X-\frac{\epsilon P}{2B}) = \int dq \: \tilde{V}(q) e^{i q (X-
\frac{\epsilon P}{2B}) } }  \end{eqnarray} where the factor ordering in the exponential is
not ambiguous since:
\begin{eqnarray} \displaystyle{
[q_i X^i, q_l \epsilon^{lj} P_j] = q_i q_l \epsilon^{lj} [X^i P_j]
= 0 } 
\label{17}
\end{eqnarray} 
Let $\langle k|$ and $|l \rangle$ be momentum eigenvectors. The commutation relations
(\ref{17}) immediately imply that
\begin{eqnarray} \displaystyle{
\langle k |\exp[iq(X-\frac{\epsilon p}{2B})] | l \rangle =
\delta(k-q-l) \exp[-i {q \epsilon l}/{2B}]}
\end{eqnarray} 
The
phase factor above is the usual Moyal bracket phase that is ubiquitous
in non-commutative geometry.

\subsection{String theory in magnetic fields}

Let us consider bosonic string theory in the presence of a
D3-brane. The coordinates of the brane are $x^0$, $x^1$, $x^2$,
$x^3$. The remaining coordinates will play no role. We will also
assume a background antisymmetric tensor field $B_{\mu \nu}$ in
the 1,2 direction. We will study the open string sector with
string ends attached to the D3-brane in the light cone frame.
\par Define
\begin{eqnarray} \displaystyle{
x^{\pm}= x^0 \pm x^3 } \end{eqnarray} and make the usual light
cone choice of world sheet time
\begin{eqnarray} \displaystyle{
\tau = x^+ } \end{eqnarray}
\par The string action is
\begin{eqnarray} \displaystyle{
{\cal L} = \frac{1}{2} \int_{-L}^L  \: d\tau \: d\sigma \: \left[
\left( \frac{\partial x^i}{\partial \tau} \right)^2 - \left(
\frac{\partial x^i}{\partial \sigma} \right)^2 + B_{ij}
\left( \frac{\partial x^i}{\partial \tau} \right) \left(
\frac{\partial x^j}{\partial \sigma} \right) \right] }
\end{eqnarray} We have numerically fixed $\alpha'$ and the parameter $L$ can
be identified with $P_-$, the momentum conjugate to $x_-$.
\par In what follows we will be interested in deriving the limit $B \longrightarrow
\infty$ of the action above, by keeping fixed the following rescaled variables
\begin{eqnarray} \displaystyle{
\left\{
\begin{array}{l}
\displaystyle{x^i= \frac{y^i}{\sqrt{B}}} \\ \tau = tB
\end{array}
\right. } \end{eqnarray}
After dropping a term which vanishes in the large $B$ limit and an integration by parts, the
action reduces to
\begin{eqnarray} \displaystyle{
{\cal L}= \frac{1}{2} \int d\sigma \; d\tau \left( \frac{\partial
y}{\partial \sigma} \right)^2 + \left. \epsilon_{ij} \dot{y_i} y_j
\right|_{-L}^L } \end{eqnarray} Since for $\sigma \neq \pm L$ the
time derivatives of $y$ do not appear in $S$ we may trivially
integrate them out. The solution of the classical equation of
motion is
\begin{eqnarray} \displaystyle{
y(\sigma) = y + \frac{ \Delta \sigma}{L} } \end{eqnarray}
with $\Delta$ and $y$ independent of $\sigma$. The resulting action is
\begin{eqnarray} \displaystyle{
{\cal L}=  \left[ -\frac{2 \Delta^2}{L} + \dot{y} \epsilon \Delta
\right] } \label{29}
\end{eqnarray} Evidently, the action is of the same form
as the model in section 1 with $B$ and $K$ rescaled. 

\subsubsection{The interaction vertex}
Interactions in light cone string theory are represented by string
splitting and joining. Consider two incoming strings with momenta
$p_1$, $p_2$ and center of mass positions $y_1$, $y_2$. If their
endpoints coincide they can join to form a third string with
momentum $-p_3$. The constraints on the endpoints are summarized
by the overlap $\delta$ function
\begin{eqnarray} &&\!\!\!\!\!\!\displaystyle{
\mathcal{V} = \delta((y_1 -  \Delta_1)-(y_2+ \Delta_2))  \: 
 \delta((y_2- \Delta_2) - (y_3 +  \Delta_3))  \: 
 \delta((y_3 -\Delta_3) - (y_1 +\Delta_1))
 }\nonumber\\
&&\label{30}
\end{eqnarray}
\par From eq.~(\ref{29}) we see that the center of mass momentum is
related to $\Delta$ by
\begin{eqnarray} \displaystyle{
P=\epsilon \Delta } \end{eqnarray} Inserting this in eq.~(\ref{30}) 
gives the vertex
\begin{eqnarray}
&&\!\!\!\!\displaystyle{
\mathcal{V} = \delta(y_1-y_2+ (\epsilon p_1 +\epsilon
p_2)) \:  \delta(y_2-y_3+ (\epsilon p_2 +\epsilon p_3)) \: 
\delta(y_3-y_1+ (\epsilon p_3 +\epsilon p_1)) }\nonumber\\
&&\end{eqnarray}
\par To get the vertex in momentum space multiply by $\displaystyle{
e^{i(p_1 y_1 + p_2 y_2 + p_3 y_3)} }$ and integrate over $y$. This
yields
\begin{eqnarray} \label{c30}  \displaystyle{
\mathcal{V}= e^{i (p_1 \epsilon p_2) } \delta(p_1 +p_2
+p_3) } \end{eqnarray} This is the usual form of the vertex in
non-commutative field theory. We have scaled the  ``transverse''
coordinates $x^1$, $x^2$ (but not $x^0$, $x^3$) and momenta so
that the $B$ field does not appear in the vertex. If we go back to
the original units the phases in (\ref{c30}) will be proportional
to $1/B$.
\par Evidently a quantum of non-commutative Yang Mills theory may be thought
of as a
straight string connecting two opposite charges. The separation
vector $\Delta$ is perpendicular to the direction of motion $P$.
\par Now consider the geometry of the 3-body vertex. The string endpoints
$u$, $v$, $w$ define a triangle with sides
\begin{eqnarray} \displaystyle{
\begin{array}{c}
\Delta_1 = (u-v) \\ \Delta_2= (v-w) \\ \Delta_3=(w-u)
\end{array}
} \end{eqnarray} and the three momenta are perpendicular to the
corresponding $\Delta$. It is straightforward to see that the
phase
\begin{eqnarray} \displaystyle{
\epsilon_{ij} p_i q_j /B }\equiv p\wedge q \end{eqnarray} is just
the area of the triangle times $B$. In other words, it is the
magnetic flux through the triangle. Note that it can be of either
sign.
\par More generally, we may consider a Feynman tree diagram constructed
from such vertices. For example consider figure (9a). The overall
phase is the total flux through the triangles A, B and C. In fact we
can simplify this by shrinking the internal propagators to get
figure (9b). Thus the phase is the flux through a polygon formed
from the $\Delta$'s of the external lines. The phase depends only
on the momenta of the external lines and their cyclic order.

\subsection{Structure of Perturbation Theory}
In this section we will consider the effects of the Moyal phases
on the structure of Feynman amplitudes in non-commutative Yang
Mills theory. Let us first review the diagram rules for ordinary
Yang Mills theory in 't Hooft double-line representation.
\par The gauge propagator can be represented as a double line as if
the gauge boson were a quark-antiquark pair as in figure (10). Each
gluon is equipped with a pair of gauge indices $i,j$, a momentum
$p$ and a polarization $\varepsilon$ satisfying $\varepsilon \cdot p=
\varepsilon^\mu p_\mu =0$. 
\par The vertex describing 3-gauge boson interaction is shown in
figure (11). In addition to Kronecker $\delta$ for the gauge indices
and momentum $\delta$ functions the vertex contains the factor
\begin{eqnarray} \displaystyle{
(\varepsilon_1 \cdot p_3 + \varepsilon_3 \cdot p_2 + \varepsilon_2 \cdot
p_1 ) } \end{eqnarray} The factor is antisymmetric under
interchange of any pair and so it must be accompanied by an
antisymmetric function of the gauge indices. For a purely abelian
theory the vertex vanishes when symmetrized.
\par Now we add the new factor coming from the Moyal bracket. This
factor is
\begin{eqnarray} \displaystyle{
e^{i p_1 \wedge p_2} = e^{i p_2 \wedge p_3} = e^{i p_3 \wedge p_1}
} \end{eqnarray} where $p_a \wedge p_b$ indicates an antisymmetric
product
\begin{eqnarray} \displaystyle{
\begin{array}{c}
p \wedge q = p_\mu q_\nu \theta^{\mu \nu} \\ \theta^{\mu \nu} = -
\theta^{\nu \mu}
\end{array}
} \end{eqnarray} Because these factors are not symmetric under
interchange of particles, the vertex no longer vanishes when Bose
symmetrized  even for the Abelian theory.
\par The phase factors satisfy an important identity. Let us consider the
phase factors that accompany a given diagram. In fact from now on
a diagram will indicate {\bf only the phase factor} from the
product of vertices. Now consider the diagram in figure (12a). It is
given by
\begin{eqnarray} \displaystyle{
e^{i(p_1 \wedge p_2)} e^{i (p_1 +p_2) \wedge p_3} = e^{i (p_1
\wedge p_2 + p_2 \wedge p_3 + p_1 \wedge p_3)} } \end{eqnarray} On
the other hand the dual diagram figure (12b) is given by $e^{i(p_1
\wedge (p_2 + p_3) + p_2 \wedge p_3)}$. It is identical to the
previous diagram. Thus the Moyal phases satisfy  old fashioned
``channel duality''. This conclusion is also obvious from the
``flux through polygon'' construction of the previous section.
\par In what follows, a ``duality move'' will refer to a replacement of a
diagram such as in figure (12a) by the dual diagram in figure (12b).
\par Now consider any planar diagram with $L$ loops. By a series of
``duality moves'' it can be brought to the form indicated in figure (13)
consisting of a tree with $L$ simple one-loop tadpoles.
\par Let us consider the tadpole, figure (14).
The phase factor is just $e^{i q \wedge q}=1$. Thus the loop
contributes nothing to the phase and the net effect of the Moyal
factors is exactly that of the tree diagram. In fact all trees
contributing to a given topology have the same phase, which is a
function only of the external momentum. The result is that for
planar diagrams the Moyal phases do not affect the Feynman
integrations at all. In particular the planar diagrams have
exactly the same divergences as in the commutative theory.
Evidently in the large N limit non-commutative field theory =
ordinary field theory\footnote{In the case of Yang-Mills theory subtleties can
arise: for example the non-commutative $U(1)$ gauge theory is a non-trivial, non-free theory
even in the planar limit and, thus, not equivalent to its commutative counterpart \cite{armoni}.}.
\par On the other hand, divergences that occur in nonplanar diagrams
are generically regulated by the phase factors. For example consider the
nonplanar diagram in figure (15). The Moyal phase for the diagram is
\begin{eqnarray} \label{np}\displaystyle{
e^{i p \wedge q} e^{i p \wedge q} = e^{2 i p \wedge q} }
\end{eqnarray} where $p$ is the extermal momentum and $q$ the momentum circulating in the loop. 
It is not difficult to see that
such oscillating phases will regulate divergent diagrams and make them finite, 
unless the diagram contains divergent planar subdiagrams. However there is an
exception to the rule that nonplanar diagrams are finite.
If a line with a nonplanar self energy
insertion such as in figure (15) happens to have vanishing momentum
in the 1,2 plane then according to eq. (\ref{np}) its phase will vanish.
Thus it seems that
the leading  high momentum behavior of the theory is controlled by the
planar diagrams and by nonplanar diagrams at exceptional values 
of the external momenta\cite{bgi}.

An interesting question arises if we study the theory on a torus
of finite size \cite{dani}. For an ordinary local field theory high momentum
behavior basically 
corresponds to small distance behavior. For this reason we expect
the high momentum behavior on a torus to be identical to that in
infinite space once the momentum becomes much larger than the
inverse size of the torus. However in the non-commutative case the
story is more interesting. We have seen that high momentum in the
1,2 plane is associated with $large$ distances in the
perpendicular direction. Most likely this means that the finite
torus generically behaves very differently at high momentum than
the infinite plane. Indeed, as we noticed above, nonplanar diagrams
diverge for exceptional values of the external momenta. 
Since the set of exceptional momenta is of zero measure, this presumably leads to no divergence
in the self-energy in infinite space
when the external momenta in question are integrated over. The situation could be
different for compact non-commutative geometries since integrals
over momenta are replaced by sums \cite{KW}.
\par The fact that the large N limit is essentially the same
for non-commutative and ordinary Yang Mills theories implies that
in the AdS/CFT correspondence the introduction of non-commutative
geometry does not change the thermodynamics of the theory \cite{HI}. It may
also be connected to the fact that in the matrix theory
construction of Connes-Douglas-Schwartz and Douglas-Hull, the
large N limit effectively decompactifies $X^{11}$ and should
therefore eliminate dependence on the 3-form potential. However
the argument is not straightforward since in matrix theory we are
not usually in the 't Hooft limit.

\section*{Acknowledgments}
The work of D.~B. is supported by a Marie Curie fellowship, contract number HPMF-CT-1999-00372. 
I owe special thanks to Stefano Giusto for his patient typographical support. 
I would also like to thank all my friends and colleagues at
Weizmann, and in particular Micha Berkooz and Adam Schwimmer, for interesting
and stimulating discussions and for their long term support to this research project.

\begin{figure}[ht] 
\epsfxsize=88mm
\epsfbox{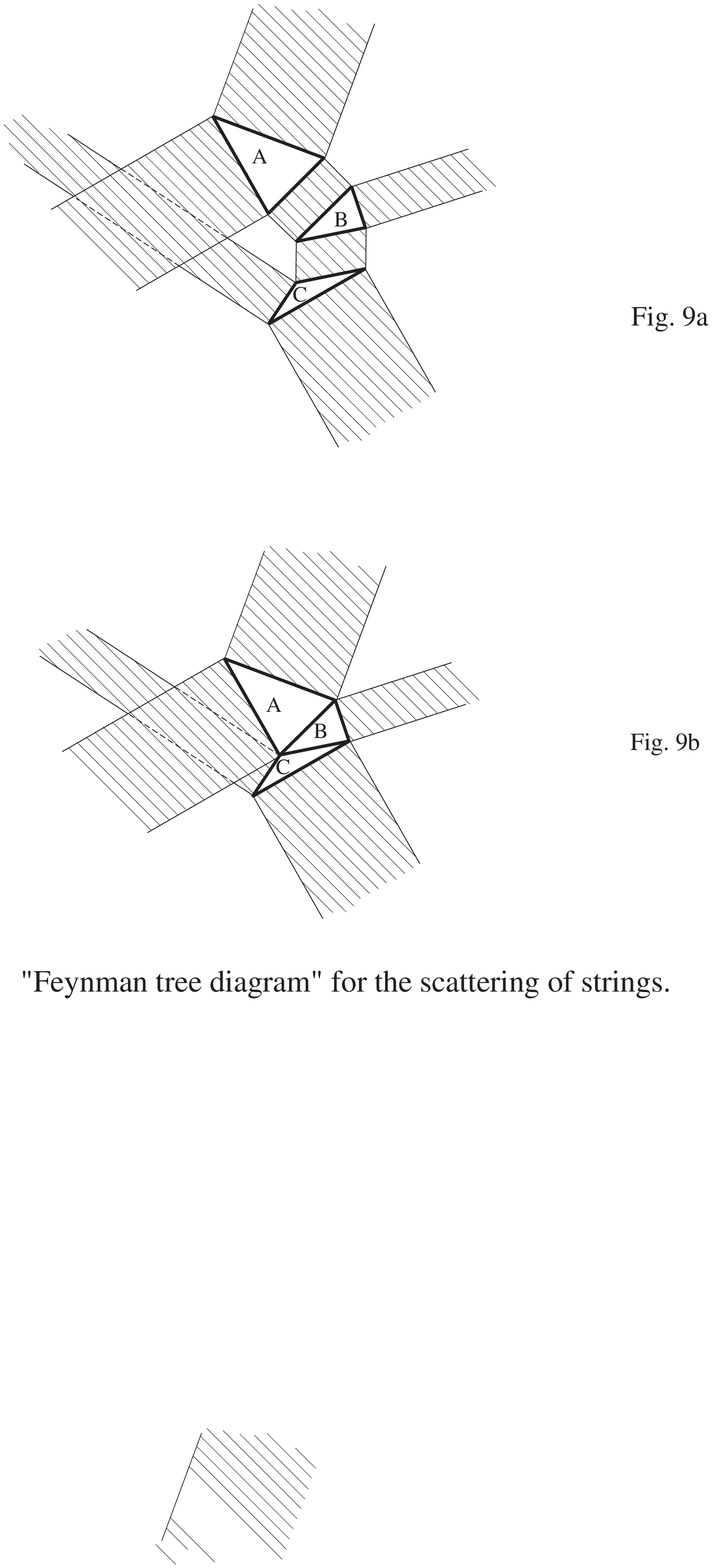}
\end{figure}

\begin{figure}[ht] 
\epsfxsize=125mm
\epsfbox{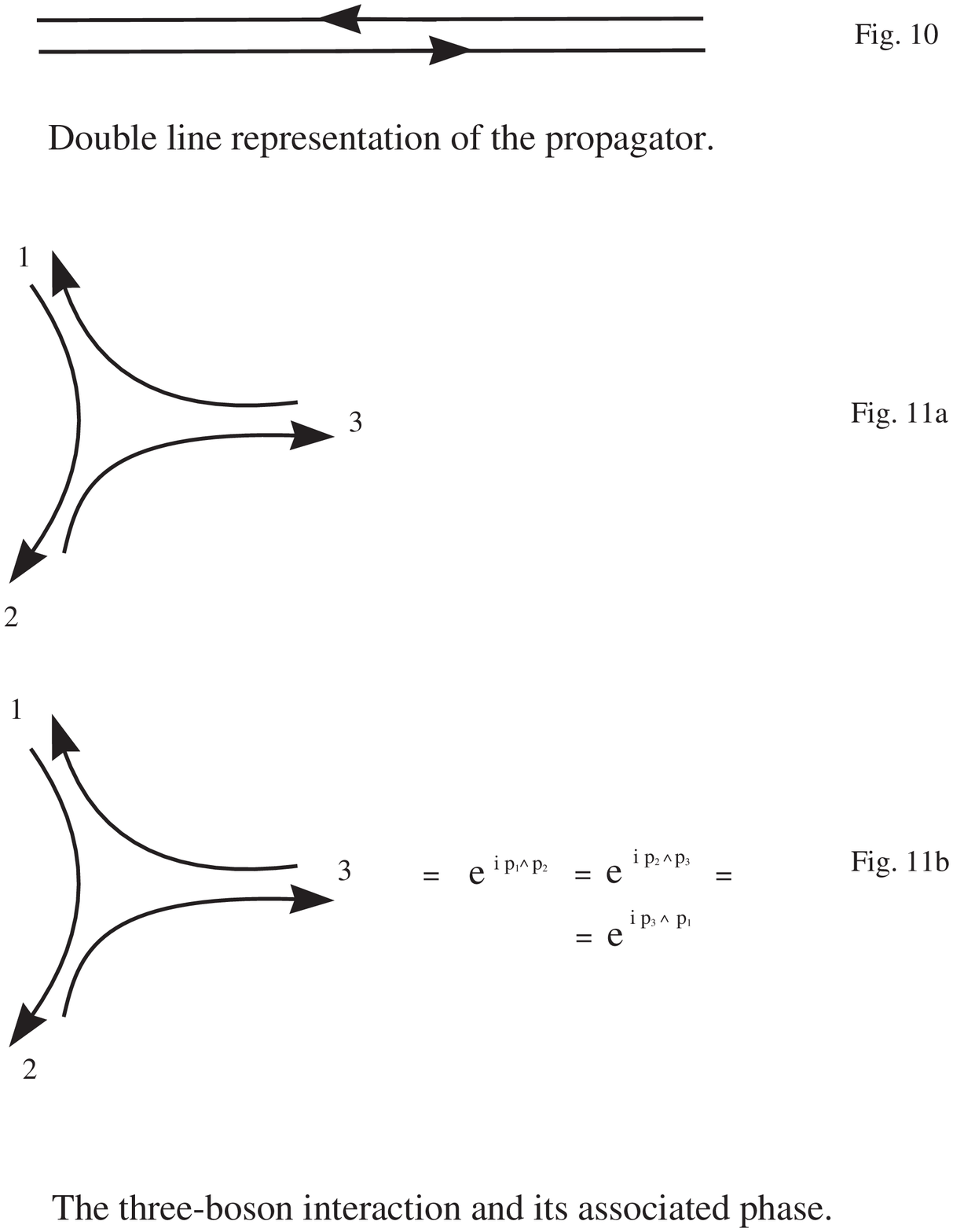}
\end{figure}

\begin{figure}[ht]
\epsfxsize=125mm
\epsfbox{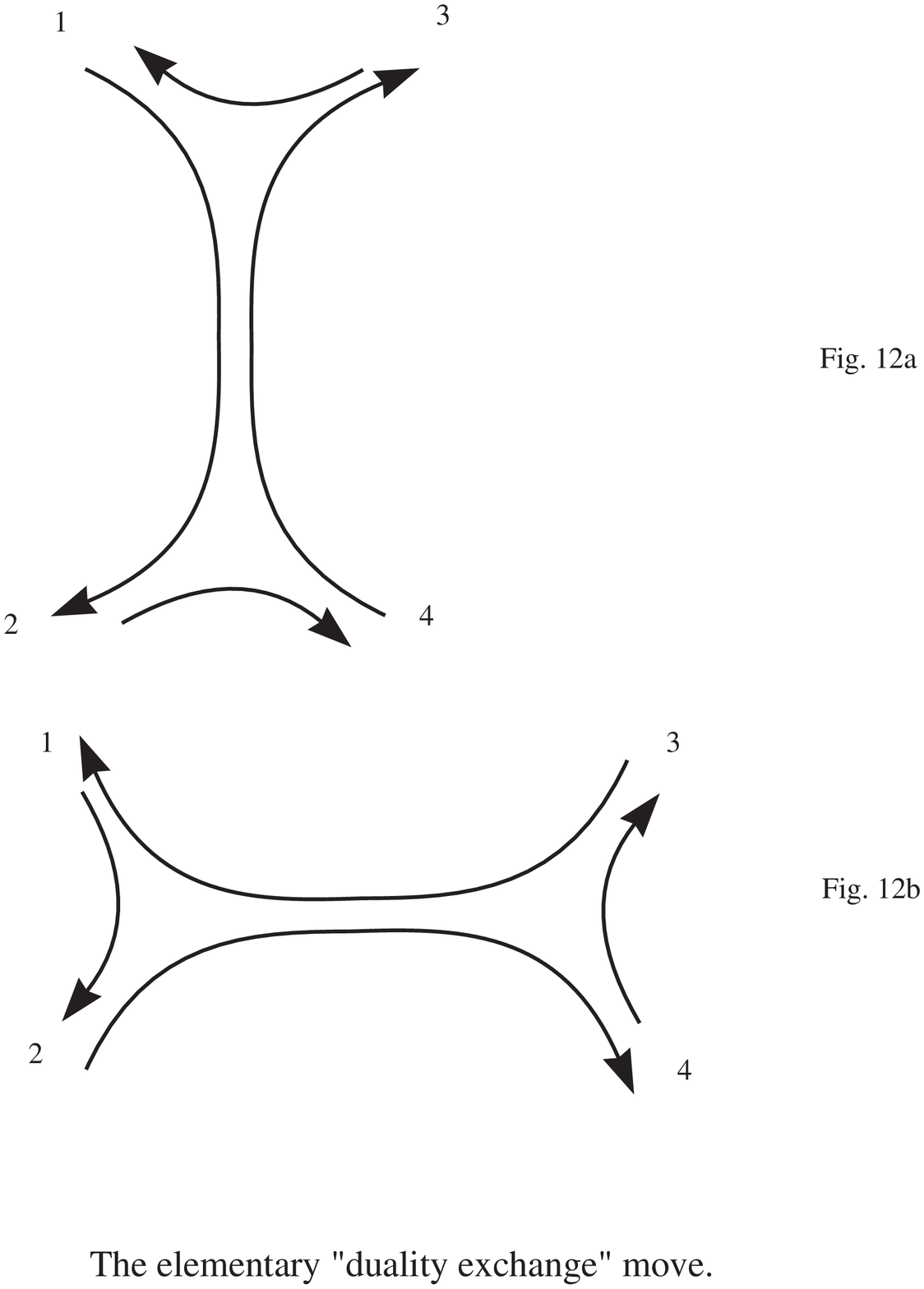}
\end{figure} 

\begin{figure}[ht]
\epsfxsize=125mm
\epsfbox{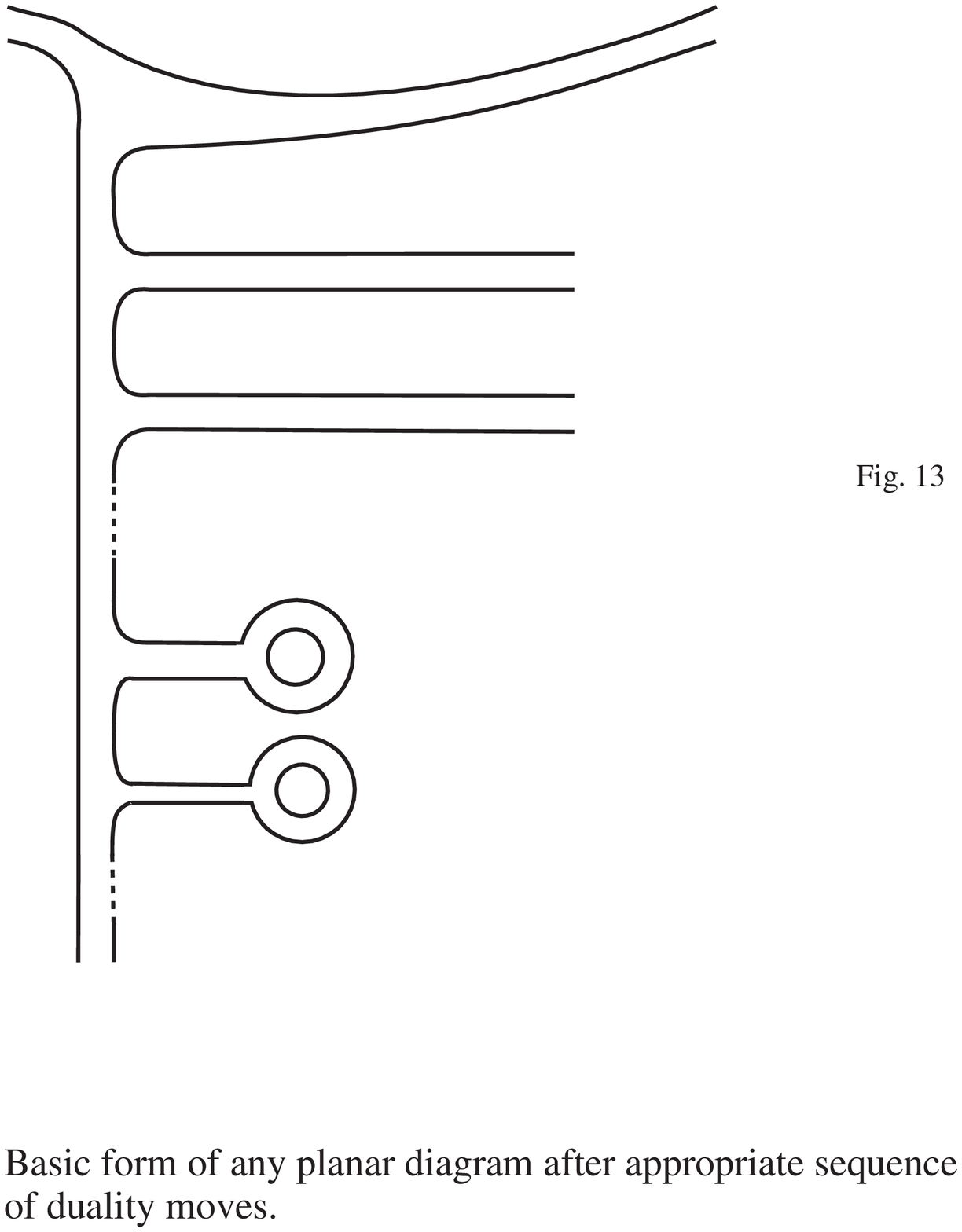}
\end{figure}

\begin{figure}[ht]
\epsfxsize=125mm
\epsfbox{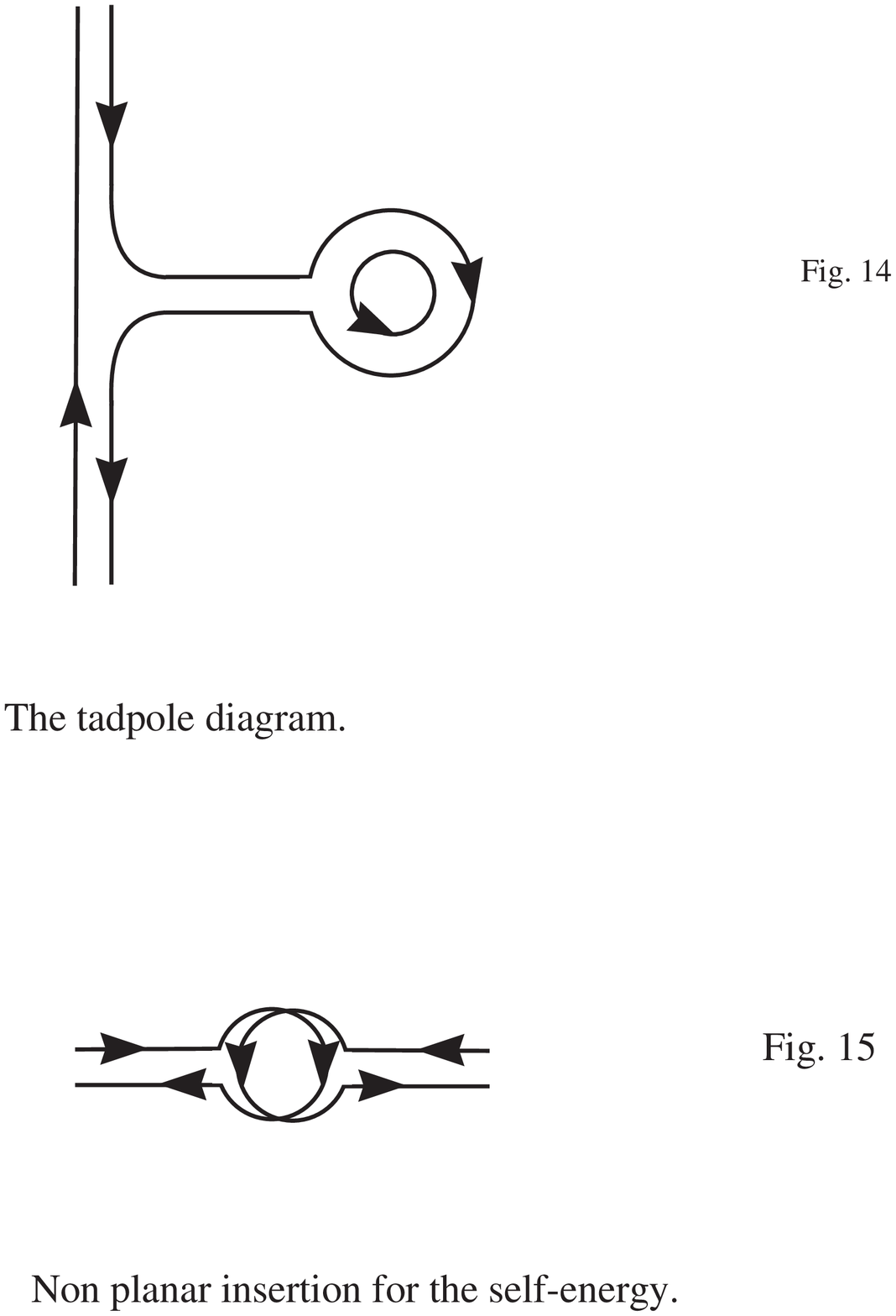}
\end{figure}

\end{document}